\documentclass[journal]{IEEEtran}

\usepackage[utf8]{inputenc}
\usepackage[T1]{fontenc}
\usepackage[scaled=0.92]{helvet}  
\usepackage[noadjust]{cite}

\usepackage{graphicx}
\usepackage{array}
\newcolumntype{M}[1]{>{\centering\arraybackslash}m{#1}}
\usepackage{tabularx,booktabs}
\newcolumntype{C}{>{\centering\arraybackslash}X} 
\usepackage{dblfloatfix}
\usepackage{amsmath}
\usepackage{newtxtext, newtxmath}
\usepackage{dsfont}
\usepackage{hhline,booktabs}
\usepackage{tablefootnote}
\usepackage{enumitem}
\usepackage[T1]{fontenc}
\usepackage{adjustbox}
\usepackage{booktabs, multirow}
\usepackage{amsmath}

\usepackage[scaled=0.9]{DejaVuSansMono}
\usepackage{listings}
\usepackage{amsfonts}

\usepackage{amsthm,thmtools,xcolor}
\declaretheoremstyle[
  headfont=\color{red}\normalfont\bfseries,
  bodyfont=\color{red}\normalfont\itshape,
]{colored}

\usepackage{bm}
\usepackage{url}
\usepackage{mathtools}
\usepackage{balance}
\usepackage{subcaption}
\allowdisplaybreaks

\usepackage{algorithm}
\usepackage{algpseudocode}
\usepackage{etoolbox}
\usepackage[mathscr]{euscript}
\usepackage{calc}
\usepackage{pgfplots}
\usepackage{tikz}
\usepackage{setspace}
\usepackage{pgfplotstable}

\usepackage[font=small]{caption}
\DeclareMathAlphabet{\pazocal}{OMS}{zplm}{m}{n}
\DeclareSymbolFont{missing}{OML}{cmr}{m}{n}
\DeclareMathSymbol{\ell}{\mathord}{missing}{'140}

\usetikzlibrary{calc}
\usetikzlibrary{patterns}
\usetikzlibrary{spy,backgrounds}
\pgfplotsset{grid style={dotted,gray}}

\newcounter{problem}
\newcounter{save@equation}
\newcounter{save@problem}

\newlength{\depthofsumsign}
\newlength{\totalheightofsumsign}
\newlength{\heightanddepthofargument}
\makeatletter
\tikzset{reset label anchor/.code={%
    \let\tikz@auto@anchor=\pgfutil@empty
    \def\tikz@anchor{#1}
  },
  reset label anchor/.default=center
}

\let\save@mathaccent\mathaccent
\newcommand*\if@single[3]{%
  \setbox0\hbox{${\mathaccent"0362{#1}}^H$}%
  \setbox2\hbox{${\mathaccent"0362{\kern0pt#1}}^H$}%
  \ifdim\ht0=\ht2 #3\else #2\fi
}
\newcommand*\rel@kern[1]{\kern#1\dimexpr\macc@kerna}
\newcommand*\widebar[1]{\@ifnextchar^{{\wide@bar{#1}{0}}}{\wide@bar{#1}{1}}}
\newcommand*\wide@bar[2]{\if@single{#1}{\wide@bar@{#1}{#2}{1}}{\wide@bar@{#1}{#2}{2}}}
\newcommand*\wide@bar@[3]{%
  \begingroup
  \def\mathaccent##1##2{%
    \let\mathaccent\save@mathaccent
    \if#32 \let\macc@nucleus\first@char \fi
    \setbox\z@\hbox{$\macc@style{\macc@nucleus}_{}$}%
    \setbox\tw@\hbox{$\macc@style{\macc@nucleus}{}_{}$}%
    \dimen@\wd\tw@
    \advance\dimen@-\wd\z@
    \divide\dimen@ 3
    \@tempdima\wd\tw@
    \advance\@tempdima-\scriptspace
    \divide\@tempdima 10
    \advance\dimen@-\@tempdima
    \ifdim\dimen@>\z@ \dimen@0pt\fi
    \rel@kern{0.6}\kern-\dimen@
    \if#31
    \overline{\rel@kern{-0.6}\kern\dimen@\macc@nucleus\rel@kern{0.4}\kern\dimen@}%
    \advance\dimen@0.4\dimexpr\macc@kerna
    \let\final@kern#2%
    \ifdim\dimen@<\z@ \let\final@kern1\fi
    \if\final@kern1 \kern-\dimen@\fi
    \else
    \overline{\rel@kern{-0.6}\kern\dimen@#1}%
    \fi
  }%
  \macc@depth\@ne
  \let\math@bgroup\@empty \let\math@egroup\macc@set@skewchar
  \mathsurround\z@ \frozen@everymath{\mathgroup\macc@group\relax}%
  \macc@set@skewchar\relax
  \let\mathaccentV\macc@nested@a
  \if#31
  \macc@nested@a\relax111{#1}%
  \else
  \def\gobble@till@marker##1\endmarker{}%
  \futurelet\first@char\gobble@till@marker#1\endmarker
  \ifcat\noexpand\first@char A\else
  \def\first@char{}%
  \fi
  \macc@nested@a\relax111{\first@char}%
  \fi
  \endgroup
}

\def\endthebibliography{%
  \def\@noitemerr{\@latex@warning{Empty `thebibliography' environment}}%
  \endlist
}

\algnewcommand{\LineComment}[1]{\Statex \hskip\ALG@thistlm
  \(//\) #1}
\def\BState{\State\hskip-\ALG@thistlm}

\algdef{SE}[DOWHILE]{Do}{EndDoWhile}{\algorithmicdo}[1]{\algorithmicwhile\ #1}%

   \tikzset{nomorepostaction/.code=\let\tikz@postactions\pgfutil@empty}

   \long\def\ifnodedefined#1#2#3{%
   \@ifundefined{pgf@sh@ns@#1}{#3}{#2}%
 }
\usetikzlibrary{decorations.pathmorphing,arrows.meta,positioning}
\tikzstyle{printersafe}=[decoration={amplitude=0pt}]

\def\ps@IEEEtitlepagestyle{
  \def\@oddfoot{\mycopyrightnotice}
  \def\@evenfoot{}
}
\def\mycopyrightnotice{
  {\footnotesize
  \begin{minipage}{\textwidth}
  \centering
Copyright (c) 2020 IEEE. Personal use of this material is permitted. However, permission to use this material for any other purposes must be obtained from the IEEE by sending a request to pubs-permissions@ieee.org.
  \end{minipage}
  }
}

\makeatother

\usetikzlibrary{automata}
\usetikzlibrary{calc,fit,shapes.geometric}
\pgfdeclarelayer{signal}
\pgfsetlayers{signal,main}
\tikzstyle{printersafe}=[segment amplitude=0 pt]
\usetikzlibrary{arrows}
\newcounter{cntr}
\usetikzlibrary{calc}
\usetikzlibrary{decorations.pathreplacing,decorations.markings,shapes.geometric}
\tikzset{naming/.style={align=center}}
\tikzset{antenna/.style={insert path={-- coordinate (ant#1) ++(0,0.5) -- +(135:0.5) + (0,0) -- +(45:0.5)}}}
\tikzset{station/.style={naming,draw,shape=dart,shape border rotate=90, minimum width=20mm, minimum height=20mm,outer sep=0pt,inner
    sep=3pt}}
\tikzset{mobile/.style={naming,draw,shape=rectangle,minimum width=12mm,minimum height=6mm, outer sep=0pt,inner sep=3pt}}
\tikzset{radiation/.style={{decorate,decoration={expanding waves,angle=90,segment length=6pt}}}}

\tikzset{
  every pin/.style={rectangle,rounded corners=3pt,font=\footnotesize},
  small dot/.style={fill=black,circle,scale=0.5}
}

\tikzset{
  invisible/.style={opacity=0},
  visible on/.style={alt={#1{}{invisible}}},
  alt/.code args={<#1>#2#3}{%
    \alt<#1>{\pgfkeysalso{#2}}{\pgfkeysalso{#3}} 
  },
}

\tikzset{pics/.cd,
  SBS/.style={code={
      \begin{scope}[local bounding box=#1]
        \fill [pic actions/.try] (-1,0) -- (-1/2,3) -- (1/2, 3) -- (1,0) -- cycle;
        \fill [pic actions/.try] (-1/16,2) rectangle (1/16,4);
        \fill [pic actions/.try] (0,4) circle [radius=1/4];
        \foreach \i in {-1,1}
        \fill [shift=(90:4), xscale=\i]
        \foreach \r in {1,3/2,2}{
          (-45:\r) arc (-45:45:\r) -- (45:\r-1/10)
          arc(45:-45:\r-1/10) -- cycle
        };
      \end{scope}
    }},
  MBS/.style={code={
      \begin{scope}[local bounding box=#1]
        \fill [pic actions/.try] (-1,0) -- (-1/2,3) -- (1/2, 3) -- (1,0) -- cycle;
        \fill [pic actions/.try] (-1/16,2) rectangle (1/16,4);
        \fill [pic actions/.try] (0,4) circle [radius=1/4];
        \foreach \i in {-1,1}
        \fill [shift=(90:4), xscale=\i]
        \foreach \r in {1,3/2,2}{
          (-45:\r) arc (-45:45:\r) -- (45:\r-1/10)
          arc(45:-45:\r-1/10) -- cycle
        };
      \end{scope}
    }},
  SU/.style={code={
      \begin{scope}[local bounding box=#1]
        \fill [even odd rule, pic actions/.try]
        (-1,-5/2) -- (-1,-1/8) -- (1,-1/8) -- (1,-5/2)
        arc (360:180:1 and 1/4) -- cycle (-1,5/2) -- (-1,1/8) -- (1,1/8) -- (1,5/2)
        arc (0:180:1 and 1/4) -- cycle (-3/4, 9/4) -- (-3/4, 3/8) -- (3/4, 3/8) -- (3/4, 9/4)
        arc (0:180:3/4 and 1/8)-- cycle
        \foreach \i in {-1,0,1}{\foreach \j in {1,2,3}{
            (-\i*1/2-3/16,-\j/2-3/4) rectangle ++(3/8, 3/8)
          }
        }
        (-1/2,-3/4) rectangle (1/2, -1/4);
      \end{scope}
    }},
  MU/.style={code={
      \begin{scope}[local bounding box=#1]
        \fill [even odd rule, pic actions/.try]
        (-1,-5/2) -- (-1,-1/8) -- (1,-1/8) -- (1,-5/2)
        arc (360:180:1 and 1/4) -- cycle (-1,5/2) -- (-1,1/8) -- (1,1/8) -- (1,5/2)
        arc (0:180:1 and 1/4) -- cycle (-3/4, 9/4) -- (-3/4, 3/8) -- (3/4, 3/8) -- (3/4, 9/4) arc (0:180:3/4 and 1/8)-- cycle
        \foreach \i in {-1,0,1}{
          \foreach \j in {1,2,3}{
            (-\i*1/2-3/16,-\j/2-3/4) rectangle ++(3/8, 3/8)
          }
        }
        (-1/2,-3/4) rectangle (1/2, -1/4);
      \end{scope}
    }},
  SIGNAL/.style={code={
      \begin{scope}[local bounding box=#1]
        \fill [pic actions/.try]
        (0,-3) -- (-1,1/2) -- (1/8,1/4) -- (0,3) -- (1,-1/2) -- (-1/8,-1/4) -- cycle;
      \end{scope}
    }},
  queuei/.style={code={
      \begin{scope}
        \stepcounter{cntr}
        \node[inner sep=0pt, outer sep=0pt,draw,rectangle split,rectangle split horizontal,minimum height=0.5cm,rectangle split parts=3]
        (queue-\thecntr) [pic actions] {};
        \draw
        (queue-\thecntr.north west) -- ++(-0.2cm,0)
        (queue-\thecntr.south west) -- ++(-0.2cm,0);
        \node[above] at ([xshift=-0.5cm]queue-\thecntr.north)
        {$Q_#1$};
      \end{scope}
    }}
}
\colorlet{sky blue}{blue!60!cyan!75!black}
\colorlet{dark blue}{blue!50!cyan}
\colorlet{chameleon}{olive!75!green}
\tikzset{signal/.style={->,draw=black, line width=0.05em, dashed,printersafe}}

\newsavebox{\mybox}
\pgfplotsset{compat=1.16}
\begin{document}

\title{Network Slicing with MEC and Deep Reinforcement Learning for the Internet of Vehicles}

\author{Zoubeir~Mlika,~\IEEEmembership{Member,~IEEE}, and Soumaya~Cherkaoui,~\IEEEmembership{Senior~Member,~IEEE}\thanks{Zoubeir Mlika and Soumaya Cherkaoui are with the research laboratory on intelligent engineering for communications and networking (INTERLAB), Faculty of Engineering, Department of Electrical and Computer Science Engineering, University of Sherbrooke, Sherbrooke J1K 2R1, Quebec, Canada, (e-mail: zoubeir.mlika@usherbrooke.ca, soumaya.cherkaoui@usherbrooke.ca).}}%

\maketitle

\begin{abstract}
  The interconnection of vehicles in the future fifth generation (5G) wireless ecosystem forms the so-called Internet of vehicles (IoV). IoV offers new kinds of applications requiring delay-sensitive, compute-intensive and bandwidth-hungry services. Mobile edge computing (MEC) and network slicing (NS) are two of the key enabler technologies in 5G networks that can be used to optimize the allocation of the network resources and guarantee the diverse requirements of IoV applications.

  As traditional model-based optimization techniques generally end up with NP-hard and strongly non-convex and non-linear mathematical programming formulations, in this paper, we introduce a model-free approach based on deep reinforcement learning (DRL) to solve the resource allocation problem in MEC-enabled IoV network based on network slicing. Furthermore, the solution uses non-orthogonal multiple access (NOMA) to enable a better exploitation of the scarce channel resources. The considered problem addresses jointly the channel and power allocation, the slice selection and the vehicles selection (vehicles grouping). We model the problem as a single-agent Markov decision process. Then, we solve it using DRL using the well-known DQL algorithm. We show that our approach is robust and effective under different network conditions compared to benchmark solutions.
\end{abstract}

\newcommand{\describeContent}[1]{%
\begingroup%
\let\thefootnote\relax%
\footnotetext{#1}%
\endgroup%
}

\IEEEpeerreviewmaketitle


\section{Introduction}\label{sec:intro}
The Internet of vehicles (IoV) is an emerging concept that enhances the existing capabilities of vehicular communication by integrating with the Internet of things (IoT). IoV is a key use-case in the upcoming beyond fifth generation (5G) wireless networks~\cite{triwinarko2021phy,alalewi20215g}. IoV creates diverse new applications with extremely diverse service requirements including ultra-high reliable and delay-sensitive, bandwidth-hungry as well as compute-intensive applications~\cite{8644392}. For example, accident reports require ultra-reliable and extremely low latency whereas high definition map sharing require high bandwidth. An important open question in today's IoV networks is ``how to support, using a unified air interface, future IoV services while guaranteeing their extremely diverse performance requirements?'' Network slicing (NS) is a potential solution to respond to this question~\cite{mlika2021network,8459911,doi:10.1002/ett.3652}. NS is a tool that enables network operators to support virtualized end-to-end networks that belongs to the principle of software defined networking~\cite{azizian2017vehicle}. It mainly allows creating different logical networks on the top of a common and programmable physical infrastructure. Another technology, namely mobile edge computing, or better known as multi-access edge computing (MEC), is considered as an important building block in the future IoV ecosystem. The joint implementation of NS and MEC is a key enabler for IoV networks. These two technologies can be used not only to guarantee the diverse requirements of IoV applications but also to deploy the diverse vehicular services at the appropriate locations~\cite{8644392}.

Optimal resource allocation in IoV would go through traditional model-based optimization techniques. Due to the complex and highly dynamic nature of IoV, such a model-based approach is not very appealing. In fact, such approach ends up with strongly non-convex optimization problems that are generally NP-hard~\cite{8943940}. Thus, a model-free machine learning approach is crucial. 

Reinforcement learning (RL) is a useful technique in solving NP-hard optimization problems. It has been applied successfully to solve very hard problems in different research areas including wireless networks~\cite{abouaomar2021service}. It is based on Markov decision process (MDP) modeling where agents learn to select the best actions through repeated interactions with an unknown environment by receiving numerical reward signals~\cite{8943940}. Deep RL (DRL) uses the strong ability of neural networks to generalize across enormous state spaces and reduce the complexity of a solution, thus improving the learning process.

\begin{figure*}
  \centering
  \includegraphics[width=\textwidth,height=13cm]{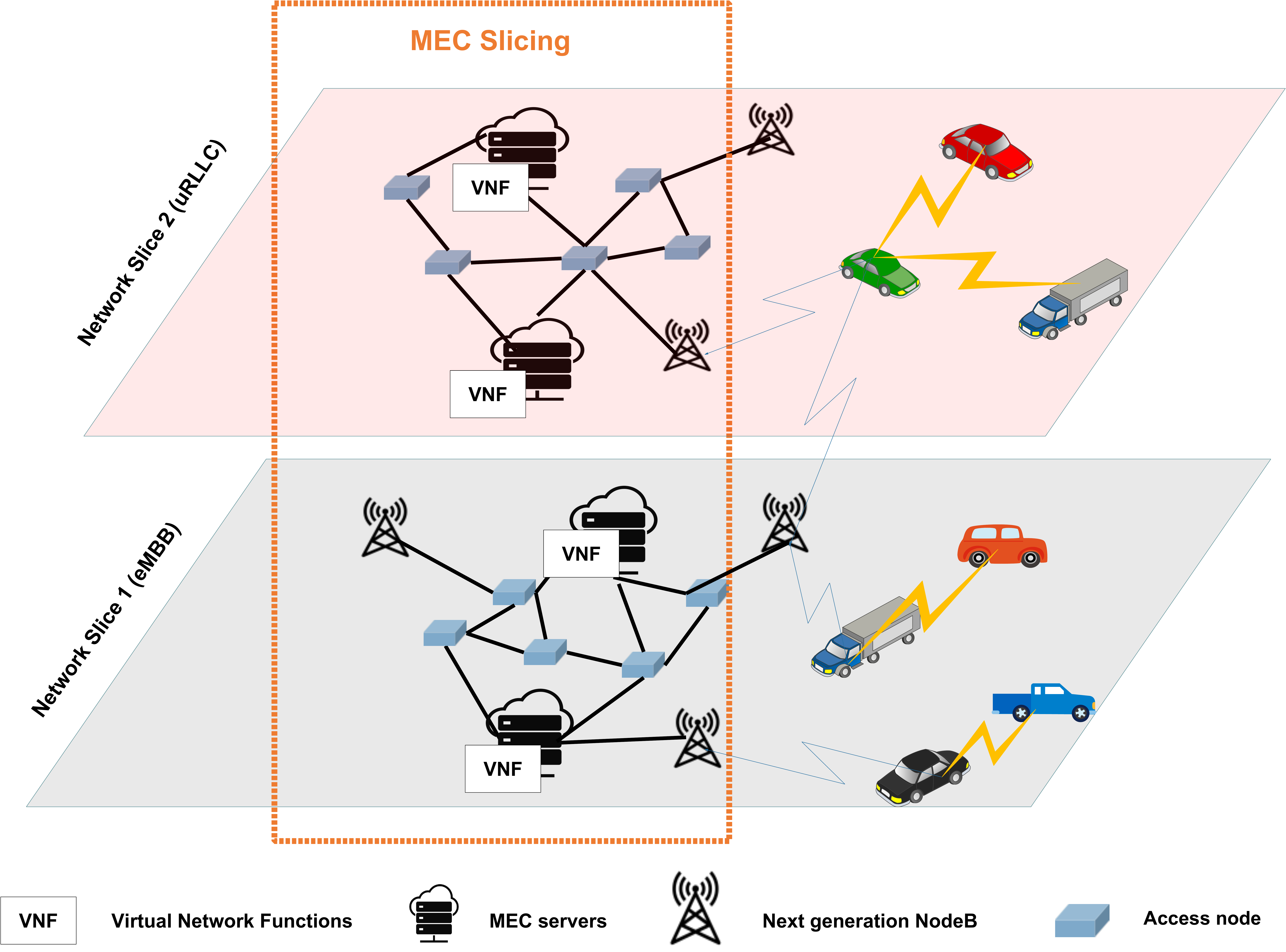}
\caption{Two network slices in an IoV-based MEC network.}
\label{sysmod1}
\end{figure*}

In this paper, using DRL, we propose a new solution framework to solve the challenging problem of resource allocation in a MEC-enabled IoV network. More specifically, we focus on the in-coverage scenario of 5G-new radio (5G-NR) in which vehicles communicate with each other through a base station, e.g., NodeB (gNB), that performs MEC-based tasks~\cite{3gpp.38.885}. We focus on the broadcast communication technique. Due to the scarce spectrum resources, non-orthogonal multiple access (NOMA) is also used in our proposed framework. NOMA is a promising technique to increase the spectral efficiency in vehicular networks~\cite{8246842}.

In more detail, the considered resource allocation problem, called IoV resource allocation (IoVRA), involves the allocation of four resources: the slice (deciding which packet to send), the coverage of the broadcast (deciding the range of the broadcast), the resource blocks (RBs), and the power. By carefully allocating these four resources, and by applying the successive interference cancellation (SIC) at the corresponding destination vehicles, NOMA can help in boosting the capacity of the IoV network. The use of NOMA in broadcast communications is different from the usual uplink and downlink NOMA techniques, which is due from the broadcast nature in IoV networks, i.e., two source vehicles broadcast with two distinct transmission powers to the same group of destination vehicles. 

Even though we propose a MEC-based IoV solution for the case of vehicle-to-vehicle (V2V) communications, our proposed system model is valid for vehicle-to-infrastructure (V2I) communications as well. Indeed, in V2I communications, a vehicle communicates with a gNB-type road side unit (RSU) or a user-type RSU through the cellular Uu or the sidelink (SL) connectivity~\cite{5gcar}. For the case of user-type RSU communications, the coverage range selection decision will simply include the RSU. For the case of gNB-type RSU communications, the broadcast coverage range selection could be ignored and replaced by RSU association. Thus, our proposed solution framework is still valid for both V2V and V2I communications.

To the best of our knowledge, this is the first work that proposes a model-free DRL framework to solve IoVRA in MEC-enabled IoV networks based on broadcast, NS and NOMA. The contributions of our work are the following. We model IoVRA as a single agent MDP. Next, we propose a deep-Q-learning (DQL) algorithm to solve it. Finally, we show that our proposed DQL algorithm outperforms benchmark algorithms.

\subsection{Organization}
The article is organized as follows. Section~\ref{sec:mod} presents the system model, the single agent MDP, and describes the proposed DQL algorithm. Section~\ref{sec:sim} presents benchmark algorithmic solutions and gives the simulation results. Finally, section~\ref{sec:cl} draws some conclusions and discusses interesting open research questions.


\section{Proposed DRL for Internet of Vehicles}\label{sec:mod}
\subsection{Internet of Vehicles Model}
We consider an IoV network composed of a set of source vehicles that generate packets, and a set of destination vehicles that receive packets. All vehicles operate in the in-coverage scenario of 5G-NR~\cite{3gpp.38.885} and thus they are covered by some gNB that performs edge computing. A source vehicle uses broadcast communications to transmit to a subset of the destination vehicles. The time is slotted into a set of slots. The total bandwidth is divided into a set of frequency slots. A resource block (RB) is given by the pair (frequency, slot).

The proposed system model supports several use cases, including advanced driving with trajectory sharing, extended sensors~\cite{9088326} and is valid for both V2V and V2I communications. To provide guaranteed quality of service requirements to the different use cases, NS is used, which is an efficient solution in IoV networks~\cite{doi:10.1002/ett.3652}. It mainly creates logical networks on the top of a common and programmable MEC-enabled IoV infrastructure. We create two network slices. The first slice (slice 1) is designed for non-safety applications such as video streaming. The second slice (slice 2) is designed for safety applications such as emergency warnings. An example of the MEC-enabled NS system model is given in Fig.~\ref{sysmod1}, where vehicles communicate with gNBs that are connected to MEC servers. On top of this network infrastructure, two network slices are created to support IoV applications. Slice 1 is designated for high throughput or enhanced mobile broadband communication (eMBB) and slice 2 is designated for ultra-reliable and low latency communication (uRLLC).

\begin{figure*}
    \centering
    \includegraphics[width=\textwidth,height=11cm]{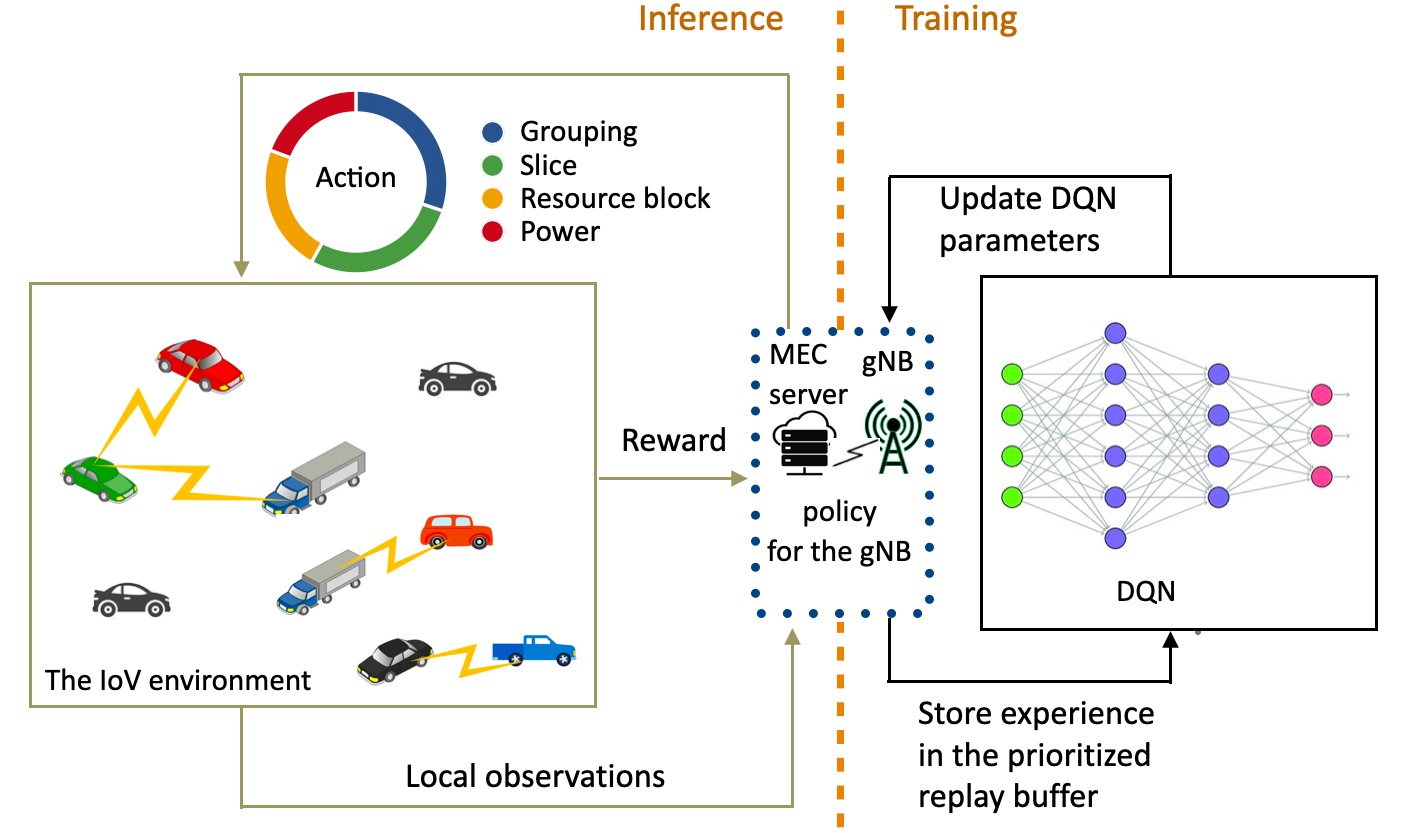} 
    \caption{IoV-based DRL architecture.}
    \label{fig:x}
\end{figure*}

Each source vehicle has two different packets for each slice, where slice 1's packet ($\textsf{pkt}^n$) requires high throughput whereas slice 2's packet ($\textsf{pkt}^s$) has stringent latency requirements. For any packet to be delivered successfully, the corresponding source vehicle requires a set of RBs such that the achievable data rates are above the minimum requirements. Packet $\textsf{pkt}^n$ can be transmitted using any RBs from the frequency-slot resource pool with a carefully chosen transmission power per each RB. However, $\textsf{pkt}^{s}$, having an arrival time and a deadline, can be transmitted using any frequency slot but only using slots between its arrival time and deadline with a carefully chosen transmission power per each RB. The wireless channel gain between two vehicles includes fast and slow fading.

A source vehicle has to decide which packet to send, at what range to broadcast, what RBs to use, and what transmission powers to allocate. The range broadcasting optimzation is smilar to the classical vehicle clustering~\cite{azizian2016distributed,azizian2016distributed2,azizian2016dcev,azizian2016optimized}. To improve the spectral efficiency of the IoV network, we use NOMA to superimpose the transmissions of the source vehicles transmitting to some destination vehicle, which uses SIC to decode the superimposed transmissions.

\subsection{Proposed Deep-Q-Learning Algorithm}
Vehicles operate in the coverage of gNB with MEC, that collects information about vehicles and performs pilot estimation to obtain the channel statistics. Based on the obtained feedback information, gNB observes the IoV environment and makes decisions. It plays the role of an intelligent entity in a single agent MDP. With the help of DRL, gNB learns to solve efficiently the complicated IoVRA problem. Specifically, gNB implements the well-known DQL approach~\cite{mnih-dqn-2015}. DQL has mainly two parts: \textit{training} and \textit{inference}. In training, gNB trains a deep-Q-network (DQN), whereas in inference, it takes actions according to its trained DQN. DQL is an improvement of the so-called QL algorithm that is based on a tabular method which creates a table of state-action pairs. QL explores the action space using an exploration policy, e.g., $\epsilon$-greedy. Despite the proven effectiveness of QL, it generally fails when the state and action spaces become large as in IoVRA.

DQL is a promising technique that is proposed to solve the curse of dimensionality in RL by approximating the Q action-value function using deep learning. One way to solve IoVRA is through multi-agent DRL by combining independent QL for each agent. That is, each agent tries to learn its own policy based on its own observations and actions while treating all other agents as part of the environment. This badly influences the result of the training as it creates a non-stationary environment that changes as other agents take decisions. For this reason, a MEC-enabled IoV network facilitates the training in such situation by modeling IoVRA as a single agent who performs the training at the edge of the IoV network. The system architecture of the proposed DQN approach is given in Fig.~\ref{fig:x}, in which gNB and MEC server interact with the IoV environment and take decisions accordingly.

Before describing in detail DQL, first, IoVRA is modeled as a single agent MDP given by the quadruple: state space, action space, reward function and transition probability. The agent in this MDP is the gNB, which takes an action, receives a reward and moves to the next state based on its interaction with the unknown IoV environment. This interaction helps gNB gain more experiences and improves its accumulated reward. 

\subsubsection{The State Space}
At any slot, any state of the IoV environment is unknown directly to gNB. Instead, gNB receives an observation from the IoV environment. In our model, an observation includes local channel state information (CSI) and the transmission behavior of the source vehicles. More precisely, an observation includes the large and small-scale fading values between vehicles. These values can be accurately estimated by the destination vehicles and fed back to gNB without significant delay~\cite{8792117}. The observation also includes a decision variable that indicates whether the source vehicles transmitted in previous slots and if so which packet did they transmit. The third observation indicates the number of leftover bits of packets that each source vehicle needs to send (e.g., initially, the number of leftover bits correspond to the packets sizes). The fourth observation element includes the arrival time and the deadline of slice 2 packets.

\subsubsection{The Action Space}
IoVRA is solved in an online fashion where at each slot, gNB makes a decision that includes (i) the broadcast coverage range selection (ii) the slice selection (iii) the RB allocation, and (iv) the power allocation. For (i), we define a discrete set of coverage distances (including zero). Thus, if gNB chooses a coverage distance (or $0$), then it will broadcast (or does not) to all destination vehicles within the chosen coverage circle having as radius the indicated range. For (ii), we define a discrete set of packets (including the empty set) that indicates which packet gNB will decide to transmit. At each slot, each source vehicle has three possible choices: it does not transmit, it transmits a slice 1 packet, or it transmits a slice 2 packet. For (iii), the RB allocation is about choosing the frequency slot to be used in the current slot. For (iv), gNB carefully chooses the transmission power per RB. Note that continuous power allocation makes the implementation of DQL more complex and thus, to keep things simple, we use a discrete set of power levels that gNB can use. Finally, the action space of gNB is given by the Cartesian product of these four discrete sets. 

\subsubsection{The Reward Signal}
We mainly focus on maximizing the packet reception ratio (PRR)~\cite{3gpp.37.885} in IoV broadcast networks. PRR is defined in as follows: for one packet and one source vehicle, the PRR is given by the percentage of vehicles with successful reception among the total number of receptions. PRR directly relates to the number of successfully received packets. Therefore, our main goal is to maximize the later.

The reward signal at any slot is the sum of individual rewards of each source vehicle. Hence, the reward signal depends on whether each source vehicle has successfully transmitted its packet or not. Technically, since we aim to maximize the number of successfully received packets, we set the reward to one once a packet is successfully delivered and zero otherwise. However, this leads to poor design since the zero individual reward leads to no useful information for learning. Thus, we build the individual reward design based on the following. When a packet is not successfully delivered or the delivery has not been completed yet, the individual reward is set to the \textit{normalized} achievable rate between the corresponding vehicles. The normalization is used to upper-bound the reward. When the packet is successfully delivered, the individual reward is set to the chosen upper-bound. In the first case, upper-bounding the individual reward helps gNB acquire useful information for future decisions whereas in the second case, choosing the individual reward to be the upper-bound teaches gNB the best possible decisions to take in the future and helps in maximizing the number of successfully delivered packets. The achievable data rate is calculated based on the signal to interference-plus-noise ratio (\textsf{sinr}) according to uplink NOMA. The overall reward signal that gNB receives is thus the sum of individual rewards of each source vehicle. The goal of DQL is to maximize the cumulative reward over the long-run, given some initial state of the IoV environment. This cumulative reward is the sum over many time steps of the weighted rewards where the weight is proportional to some constant called the discount factor. This discount factor makes future rewards more important for gNB agent as their corresponding weight becomes larger. In IoVRA problem, since the proposed MDP model consists of episodes of finite length, i.e., each episode lasts a finite number of slots, IoVRA belongs to the finite horizon set of problems~\cite{10.5555/2815660}. Further, since we aim to maximize the number of successfully delivered packets, then the MEC-based gNB agent can simply choose the discount factor to be one or a number that is close to one in order to accumulate higher rewards and thus a higher number of successfully delivered packets.

\subsubsection{The Probability Transition}
The probability of moving to the next state while being in an old state and taking some action depends on the highly dynamic IoV environment and cannot be explicitly calculated. This transition happens due to the channel coefficients variation and vehicles mobility.

\subsubsection{Training in DQL}
The DQL algorithm is composed of two parts: \textit{training} and \textit{inference}. The training is composed of several episodes where each episode spans the number of slots. DQL uses DNNs to approximate the Q function. We leverage DQL with prioritized replay memory and dueling. In general experience replay memory helps to remember and use past experiences. Standard replay memory is used to sample experience transitions uniformly without paying attention to the significance of the sampled experiences. Prioritized experience replay memory is proposed to pay more attention to important experiences. This indeed makes the learning better. Also, dueling is proposed as a new neural network architecture that represents two estimators for the Q function.

In detail, the training lasts a number of episodes and requires as input the IoV environment which includes the vehicles, the channel coefficients, the packet requirements, the available RBs and any other relevant IoV network parameter. It returns as output the trained DQN. The first step in DQL is to start the simulator which generates the vehicles and all network parameters, then it initializes the DQN hyperparameters. In the beginning of the first slot, the initial state of the IoV environment (initial distances of the vehicles, etc.) is revealed to gNB. Next, DQL iterates the episodes. For each episode, the environment is built by (i) updating the network parameters, e.g., the leftover bits of each source vehicle are updated based on the previous episodes, and (ii) moving the vehicles according to the mobility model. Next, the exploration rate $\epsilon$ is annealed based on the episode index. Annealing the exploration rate over time is a technique used in RL to solve the dilemma between exploration and exploitation, i.e., as the time goes by, we decrease $\epsilon$ to increase the exploitation probability as the agent starts to learn something useful. After a few episodes, the value of $\epsilon$ is no longer decreased. Then, gNB chooses for each source vehicle an action that is a tuple of the coverage distance, the packet, the frequency slot, and the power level. Once gNB agent chooses its action according to the annealed $\epsilon$, it calculates the reward signal. Specifically, a destination vehicle calculates the received \textsf{sinr}, finds the number of bits a source vehicle is transmitting, and communicates this information to gNB using feedback channels. The environment moves to the next state and gNB adds to its prioritized replay memory the actual experience with some associated priority, i.e., the obtained tuple (state, action, reward, next state) is associated some priority. Initially, gNB assigns random priorities to its experiences but the priorities change as it starts to learn and updates its DQN parameters. gNB samples a mini-batch from its prioritized replay memory according to their priorities that forms a dataset used to train the DQN. gNB uses a variant of the well-known stochastic gradient descent to minimize the loss and it updates the priorities of the sampled experiences proportionally to the value of the loss. Finally, once in a while, the trained DQN is copied into the target DQN.

\subsubsection{Implementing DQL}
The inference of DQL is as follows (see Fig.~\ref{fig:x}). First, the trained DQN is loaded. Also, the annealed $\epsilon$ is loaded from the last training episode (the index of the episode is also revealed). Then, for each episode (which represents a new random channel realization), the environment is reset and built---initializing the network parameters and the transmission behaviors of each agent. Next, for each slot, gNB agent, after observing the environment, chooses the best action according to its trained DQN after feedback communication between itself and the destination vehicles. Then, the reward signal is obtained, and the next episode starts with a new random channel realization.

The inference in DQL is done in an online fashion. That is, it is executed in each slot without knowing the future observations. The training in DQL is the most computationally intensive task. It is executed for a large number of episodes and can be done in an offline manner with different channel conditions and IoV network topologies. Note that training in DQL needs to be re-executed only when the topology of the IoV network undergoes significant changes, depending on the IoV network dynamics.

\section{Performance Evaluation}\label{sec:sim}
In this section, we validate the proposed DQL method. The simulation setup is based on the highway scenario of~\cite{3gpp.37.885} and most simulation parameters are taken from~\cite{8792382,9026965}. We consider a six-lane highway with a total length of $2$ km where each lane has a width of $4$ m. There are three lanes for the forward direction (vehicles move from right to left) and three lanes for the backward direction. The source and destination vehicles are generated according to spatial Poisson process. Vehicles' speed determine the vehicle density and the average inter-vehicle distance (in the same lane) is $2.5\text{s}\times v$ where $v$ is the vehicle absolute speed. The speed of a vehicle depends on its lane: the $i$th forward lane (from top to bottom with $i\in\{1,2,3\}$) is characterized by the speed of $60+2(i-1)\times10$ km/h, whereas the $i$th backward lane (from top to bottom with $i\in\{1,2,3\}$) is characterized by the speed of $100-2(i-1)\times10$ km/h. The number of source vehicles $m$ and destination vehicles $n$ is randomly chosen. The important simulation parameters are given as follows~\cite{8792382,9026965}. The carrier frequency is $2$ GHz, the per-RB bandwidth is $1$ MHz, the vehicle antenna height is $1.5$ m, the vehicle antenna gain is $3$ dBi, the vehicle receiver noise figure is $9$ dB, the shadowing distribution is log-normal, the fast fading is Rayleigh, the pathloss model is LOS in WINNER + B1, the shadowing standard deviation is $3$ dB, and the noise power $N_0$ is $-114$ dBm.

Unless specified otherwise, the slice 1 packet's size is randomly chosen in $\{0.1..1\}$ Mb. The slice 2 packet's size is $600$ bytes. gNB chooses a coverage (in m) from the set $\{100,400,1000,1400\}\cup\{0\}$. The power levels (in dBm) are given by $\{15,23,30\}\cup\{-100\}$ where $-100$ dBm is used to indicate no transmission. We set $m=3$, $n=4$, $F=2$, and $T=20$; each slot has duration $5$ ms. The DQN is trained in the Julia programming language using Flux.jl. The DQN consists of an input and an output layer and of three fully connected hidden layers containing respectively $256$, $128$, and $120$ neurons. The ReLu activation function is used in each layer. The ADAM optimizer with a learning rate of $10^{-5}$ is used. The training lasts $3000$ episodes with an exploration rate starting from $1$ and annealed to reach $0.02$ for the $80\%$ of the episodes. 

To the best of our knowledge, there are no current research works that solve IoVRA while considering the slice selection, the broadcast coverage selection, the RBs and the power allocation. We implement three benchmarks: two are based on NOMA and one is based on OMA. The partial idea of all benchmarks comes from~\cite{8632657} which is based on the swap matching algorithm. All benchmarks are centralized in the edge and offline. They are called OMA-MP, NOMA-MP, and NOMA-RP. In OMA-MP, every RB is used by at most one vehicle and the maximum transmission power is allocated. In NOMA-MP and NOMA-RP, every RB can be shared, and the maximum transmission power or a random transmission power are allocated, respectively. The coverage and slice selections are decided randomly at the beginning of each slot. The allocation of the RBs to the vehicles is done similarly in all benchmarks. First, an initial RB allocation is executed that gives the highest sum of channel power gain between a source vehicle and its destination vehicle. Once the initial allocation is obtained, a swap matching is performed to improve the number of packets successfully received. If no swap improves the matching, then the algorithm terminates.

In the simulation results, we present two performance metrics: the cumulative rewards for training the DQL and the number of successfully received packets for the inferring DQL. In the training, the reward signal received by gNB is given by the sum of the individual rewards of each source vehicle. The individual reward is equal either to (i) the upper-bounded achievable rate or to (ii) the upper bound. The event (i) happens when a packet is not yet delivered whereas the event (ii) happens when a packet is completely and successfully delivered. In the inference, the reward signal is simply given as the total number of successfully delivered packets.

\begin{figure}[ht!]
  \centering
  \captionsetup{justification=centering,margin=2cm}
  \resizebox{.45\textwidth}{!}{%
    \begin{tikzpicture}[
    every axis/.style={
        xlabel={Episodes ($\times100$)},
        ylabel={Avg. rewards of the last 200 episodes},
        ymin=11.4,ymax=12.5,
        xticklabels={2,8,14,20,28},
        xtick={200,800,1400,2000,2800},
        xmin=200,xmax=2801,
        x label style={font=\footnotesize},
        y label style={font=\footnotesize}, 
        ticklabel style={font=\footnotesize},
    }]
    \begin{axis}
    \addplot[blue] table [x=x, y=y, col sep=comma,] {avg200rewards.csv};
    \end{axis}
    \end{tikzpicture}
  }
  \caption{Training rewards.}
  \label{fig:1a}
\end{figure}
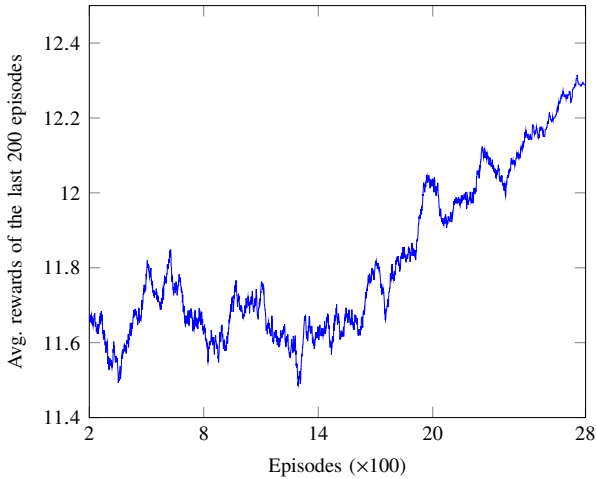
Fig.~\ref{fig:1a} illustrates the convergence of the proposed DQL algorithm versus training episodes. The figure shows the cumulative average rewards per episode where the average is taken over the last $200$ episodes. It is clear that the average reward improves as the training episodes increase. This shows the effectiveness of the proposed algorithm. The training in DQL gradually converges starting from the episode number $\approx 2700$. Note that the convergence of the algorithm is not smooth and contains some fluctuations which is due mainly to the high mobility nature of the IoV environment. Based on Fig.~\ref{fig:1a}, DQN is trained for $3000$ episodes to provide some convergence guarantees.

In the next two figures, we present, as a performance metric, the reward obtained in the inference part of DQL, which is the number of successfully received packets. We show this performance metric as stacked bars where each bar is divided into two parts: the lower part indicates the number of successfully delivered slice 1 packets and the higher part indicates the number of successfully delivered slice 2 packets.

\begin{figure}[ht!]
  \centering
  \captionsetup{justification=centering,margin=2cm}
  \resizebox{.45\textwidth}{!}{%
    \begin{tikzpicture}[
    every axis/.style={
        xlabel={Safety message sizes ($\times300$ Bytes)},
        ylabel={Avg. number of successfully received packets},
        ybar stacked,
        ymin=0,ymax=11,
        bar width=6pt,
        xtick={2,4,6,8,10},
        width=10cm,
        x label style={font=\footnotesize},
        y label style={font=\footnotesize}, 
        ticklabel style={font=\footnotesize},
    }]
    \begin{axis}[hide axis, bar shift=-12pt,legend style={at={(0,-.2)},font=\scriptsize,anchor=north,},]
    \addplot[black,fill=yellow] coordinates {
        (2, 2.03) (4, 2.03) (6, 2.03) (8, 2.03) (10, 2.03)
    };
    \addplot[black,fill=yellow,postaction={pattern=north east lines}] coordinates {
        (2, 3.69) (4, 3.67) (6, 3.65) (8, 3.63) (10, 3.61)
    };
    \legend{\strut NOMA-MP (Slice 1),\strut NOMA-MP (Slice 2)}
    \end{axis}
    \begin{axis}[hide axis, bar shift=-4pt,legend style={at={(.35,-.2)},font=\scriptsize,anchor=north,},]
    \addplot+[black,fill=violet,postaction={pattern=horizontal lines}] coordinates {
        (2, 1.28) (4, 1.28) (6, 1.28) (8, 1.28) (10, 1.28)
    };
    \addplot+[black,fill=violet,postaction={pattern=vertical lines}] coordinates {
        (2, 4.21) (4, 4.14) (6, 4.04) (8, 3.9) (10, 3.84)
    };
    \legend{\strut NOMA-RP (Slice 1),\strut NOMA-RP (Slice 2)}
    \end{axis}
    \begin{axis}[hide axis, bar shift=4pt,legend style={at={(.7,-.2)},font=\scriptsize,anchor=north,},]
    \addplot+[black,fill=magenta,postaction={pattern=grid}] coordinates {
        (2, 1.99) (4, 1.99) (6, 1.99) (8, 1.99) (10, 1.99)
    };
    \addplot+[fill=magenta,postaction={pattern=dots}] coordinates {
        (2, 3.53) (4, 3.51) (6, 3.5) (8, 3.46) (10, 3.43)
    };
    \legend{\strut OMA-MP (Slice 1),\strut OMA-MP (Slice 2)}
    \end{axis}
    \begin{axis}[bar shift=12pt,legend style={at={(1,-.2)},font=\scriptsize,anchor=north,},]
    \addplot+[black,fill=red,postaction={pattern=north west lines}] coordinates {
        (2, 1.77) (4, 1.66) (6, 1.74) (8, 1.69) (10, 1.69)
    };
    \addplot+[black,fill=red,postaction={pattern=crosshatch}] coordinates {
        (2, 7.88) (4, 7.19) (6, 6.72) (8, 6.24) (10, 5.81)
    };
    \legend{\strut DQL (Slice 1),\strut DQL (Slice 2)}
    \end{axis}
    \end{tikzpicture}
  }
  \caption{Impact of safety message sizes}
  \label{fig:1b}
\end{figure}
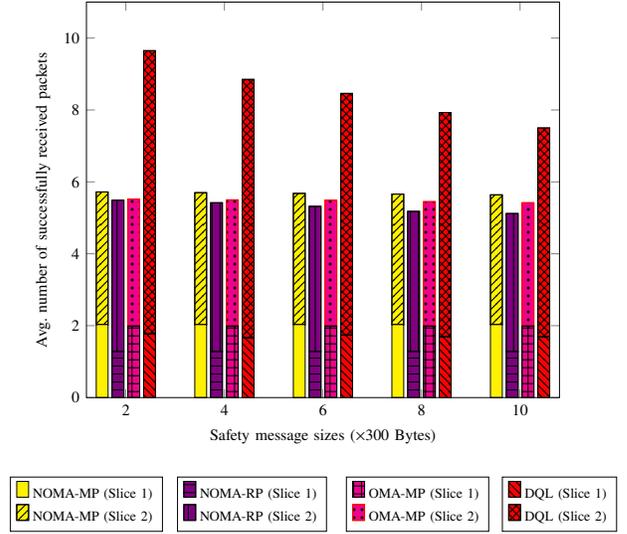
Fig.~\ref{fig:1b} shows the performance of DQL against the benchmarks when varying the slice 2 packet sizes. We can see that DQL succeeds in delivering more packets without having the full and future CSI as in the benchmarks. For example, DQL can, on average, deliver successfully almost $9$  packets. However, other benchmarks can only deliver, on average, almost $6$ packets. NOMA-RP achieves the lowest performance as expected. Further, DQL achieves a higher number of successfully delivered slice 2 packets. This is particularly important in IoV communication as slice 2 packets are mainly safety packets and thus must have a higher priority of being delivered. 

\begin{figure}[ht!]
  \centering
  \captionsetup{justification=centering,margin=2cm}
  \resizebox{.45\textwidth}{!}{%
   \begin{tikzpicture}[
    every axis/.style={
        xlabel={Safety message deadlines ($\times5$ ms)},
        ylabel={Avg. number of successfully received packets},
        ybar stacked,
        ymin=0,
        ymax=11,
        xtick={2,3,4,5,6,7,8},
        bar width=6pt,
        width=10cm,
        x label style={font=\footnotesize},
        y label style={font=\footnotesize}, 
        ticklabel style={font=\footnotesize},
    }]
    \begin{axis}[hide axis, bar shift=-12pt,legend style={at={(0,-.2)},font=\scriptsize,anchor=north,},]
    \addplot[black,fill=yellow] coordinates {
        (2,2.15) (3,2.16) (4,2.08) (5,2.63) (6,2.41) (7,2.31) (8,2.41)
    };
    \addplot[black,fill=yellow,postaction={pattern=north east lines}] coordinates {
        (2,3.26) (3,3.46) (4,3.62) (5,3.29) (6,4.11) (7,3.45) (8,3.69)
    };
    \legend{\strut NOMA-MP (Slice 1),\strut NOMA-MP (Slice 2)}
    \end{axis}
    \begin{axis}[hide axis, bar shift=-4pt,legend style={at={(0.35,-.2)},font=\scriptsize,anchor=north,},]
    \addplot+[black,fill=violet,postaction={pattern=horizontal lines}] coordinates {
        (2,2.66) (3,1.93) (4,2.01) (5,1.9) (6,2.41) (7,1.97) (8,1.52)
    };
    \addplot+[black,fill=violet,postaction={pattern=vertical lines}] coordinates {
        (2,2.14) (3,3.28) (4,3.22) (5,3.49) (6,3.2) (7,3.44) (8,4.21)
    };
    \legend{\strut NOMA-RP (Slice 1),\strut NOMA-RP (Slice 2)}
    \end{axis}
    \begin{axis}[hide axis, bar shift=4pt,legend style={at={(.7,-.2)},font=\scriptsize,anchor=north,},]
    \addplot+[black,fill=magenta,postaction={pattern={grid}}] coordinates {
        (2,2.35) (3,2.53) (4,2.33) (5,2.34) (6,2.75) (7,2.34) (8,2.31)
    };
    \addplot+[fill=magenta,postaction={pattern=dots}] coordinates {
        (2,3.25) (3,3.32) (4,3.86) (5,3.59) (6,3.46) (7,3.53) (8,3.53)
    };
    \legend{\strut OMA-MP (Slice 1),\strut OMA-MP (Slice 2)}
    \end{axis}
    \begin{axis}[bar shift=12pt,legend style={at={(1,-.2)},font=\scriptsize,anchor=north,},]
    \addplot+[black,fill=red,postaction={pattern={north west lines}}] coordinates {
        (2,2.24) (3,3.06) (4,3.13) (5,3.43) (6,3.52) (7,2.5) (8,2.43)
    };
    \addplot+[black,fill=red,postaction={pattern=crosshatch}] coordinates {
        (2,4.25) (3,5.03) (4,5.55) (5,3.54) (6,3.89) (7,7.52) (8,7.72)
    };
    \legend{\strut DQL (Slice 1),\strut DQL (Slice 2)}
    \end{axis}
    \end{tikzpicture}
  }
  \caption{Impact of safety message deadlines}
  \label{fig:1c}
\end{figure}
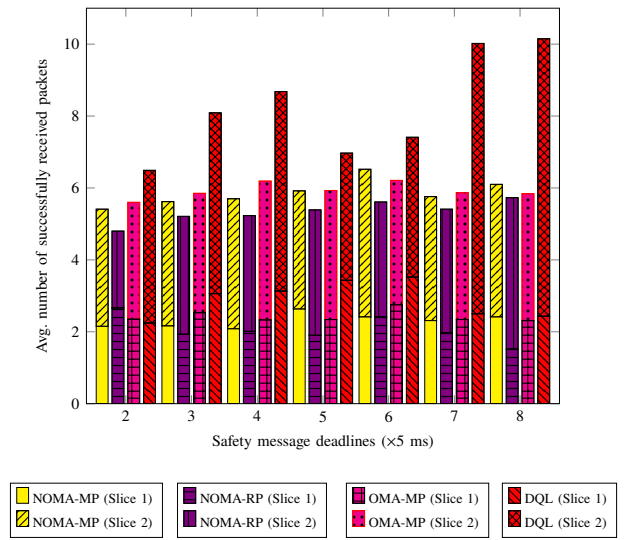
Fig.~\ref{fig:1c} shows the performance of DQL against the benchmarks when varying the slice 2 packets deadlines. DQL still achieves the best performance when the deadline of the safety packets increases. The gap between DQL and other benchmarks widens further as the deadline increases. We further notice that NOMA-RP has the worst performance for all algorithms which shows the need of a suitable power allocation method in IoVRA.  

We notice from both Fig.~\ref{fig:1b} and Fig.~\ref{fig:1c} that there is an unfair allocation of resources between the packets of the two slices. This is mainly due to highly dynamic nature of the IoV network (e.g., vehicle positions,  their speeds, etc.). For example, if a source vehicle is located close to a destination vehicle, then the quality of the wireless link between both vehicles will likely be good. Thus, gNB learns through DQL to equally likely transmit both packets. However, in the case where the source vehicle is located far away from the corresponding destination vehicle, the quality of the wireless link between both parties will probably be poor and thus, gNB will likely learn through DQL to transmit only slice 2 packets to guarantee a successful V2V communication (since slice 2 packets might not require a large number of RBs compared to slice 1 packets). It is thus important to study the fairness among different slices in such IoV network, which will be investigated in our future works.

\section{Conclusions and Future Works}\label{sec:cl}
In this paper, we developed an online MEC-based scheme to solve the slice selection, coverage selection, resource block and non-orthogonal multiple access power allocation problem in the Internet of vehicles network. We modelled the problem as a single agent Markov decision process and developed a DQL algorithm. The proposed DQL algorithm is proven robust and effective against various system parameters including the high mobility characteristics of IoV networks. It also outperformed some baseline benchmark algorithms that are based on global and offline decisions. In future works, we will investigate a two-time scale DRL approach that decides for coverage and slice selection on a slower time scale. Further, we will study the fairness of multiple slices. Finally, we will extend our system model to include mmWave communications.

\section{Acknowledgment}
The authors would like to thank the Natural Sciences and Engineering Research Council of Canada (NSERC) and the Fonds de recherche du Quebec - Nature et technologies (FRQNT),  for the financial support of this research.

\bibliographystyle{IEEEtran}
\bibliography{IEEEabrv,magasin1}
\end{document}